\documentclass[10pt,conference]{IEEEtran}
\IEEEoverridecommandlockouts
\usepackage{cite}
\usepackage{amsmath,amssymb,amsfonts}
\usepackage{algorithmic}
\usepackage{graphicx}
\usepackage{textcomp}
\usepackage{xcolor}

\usepackage{multirow}

\begin{document}

\title{Extreme mutation testing in practice:\\ An industrial case study
\thanks{This research was supported by Advantest as part of the Graduate School ``Intelligent Methods for Test and Reliability'' (GS-IMTR) at the University of Stuttgart.}
}

\author{\IEEEauthorblockN{Maik Betka}
\IEEEauthorblockA{\textit{Institute of Software Engineering} \\
\textit{University of Stuttgart}\\
Stuttgart, Germany \\
maik.betka@iste.uni-stuttgart.de}
\and
\IEEEauthorblockN{Stefan Wagner}
\IEEEauthorblockA{\textit{Institute of Software Engineering} \\
\textit{University of Stuttgart}\\
Stuttgart, Germany \\
stefan.wagner@iste.uni-stuttgart.de}
}

\maketitle

\begin{abstract}
Mutation testing is used to evaluate the effectiveness of test suites.
In recent years, a promising variation called extreme mutation testing emerged that is computationally less expensive.
It identifies methods where their functionality can be entirely removed, and the test suite would not notice it, despite having coverage.
These methods are called pseudo-tested.
In this paper, we compare the execution and analysis times for traditional and extreme mutation testing and discuss what they mean in practice.
We look at how extreme mutation testing impacts current software development practices and discuss open challenges that need to be addressed to foster industry adoption.
For that, we conducted an industrial case study consisting of running traditional and extreme mutation testing in a large software project from the semiconductor industry that is covered by a test suite of more than 11,000 unit tests.
In addition to that, we did a qualitative analysis of 25 pseudo-tested methods and interviewed two experienced developers to see how they write unit tests and gathered opinions on how useful the findings of extreme mutation testing are.
Our results include execution times, scores, numbers of executed tests and mutators, reasons why methods are pseudo-tested, and an interview summary.
We conclude that the shorter execution and analysis times are well noticeable in practice and show that extreme mutation testing supplements writing unit tests in conjunction with code coverage tools.
We propose that pseudo-tested code should be highlighted in code coverage reports and that extreme mutation testing should be performed when writing unit tests rather than in a decoupled session.
Future research should investigate how to perform extreme mutation testing while writing unit tests such that the results are available fast enough but still meaningful.
\end{abstract}

\section{Introduction}
Mutation testing is a software testing technique to measure how effective a test suite is.
Despite being around for several decades and thoroughly studied in research, it is still not widely adopted in industry \cite{jia2010analysis,papadakis2019mutation}.
Extreme mutation testing is a variation of mutation testing and was introduced by Niedermayr, Juergens, and Wagner in 2016~\cite{niedermayr2016will}.
This variation seems promising because it is computationally less expensive and easier to comprehend due to its higher abstraction level.

We want to find out how big the time improvements of extreme mutation testing over traditional mutation testing are, what they mean in practice, and which factors currently prevent industry adoption.
Therefore, we conducted an industrial case study where we performed both mutation testing techniques in a large software project from the semiconductor industry and interviewed two experienced developers about their software testing practices and the findings of extreme mutation testing in particular.

\section{Foundations and related work}
\subsection{Mutation testing}
For mutation testing, small controlled changes in the source code of the software under test are introduced to change its behavior.
The types of changes are called \textit{mutators}.
The respective altered software versions are called \textit{mutants}.
Afterwards, the given test suite is re-run.
If no test fails, this means that the test suite is not able to detect the mutant.
Thus, it is said that the mutant \textit{survived}.
Otherwise, it is said that the mutant is \textit{killed}.
Some software changes may produce syntactically equivalent mutants, i.e., mutants that do not have a different behavior.
These mutants are called \textit{equivalent mutants}.
The goal is to strengthen the test suite by enhancing or writing new tests to kill the surviving non-equivalent mutants.
The ratio of killed mutants to all generated non-equivalent mutants is called the \textit{mutation score}.
Thus, the goal of mutation testing is typically expressed by optimizing the mutation score towards one \cite{jia2010analysis,papadakis2019mutation}.

\subsection{Extreme mutation testing}
In contrast to traditional mutation testing, where mutations are usually performed on an instruction level, extreme mutation testing performs these mutations on a method level.
The goal is to find \textit{pseudo-tested methods}, which are methods where their whole functionality can be removed, and still, no test will fail.~\cite{niedermayr2016will}.
Pseudo-tested methods are surprisingly common.
In another study, 19 open-source projects have been analyzed and the median proportion of pseudo-tested methods to the total number of mutated methods was 10.1\%~\cite{niedermayr_evaluation_2019}.

Particular mutants have to survive to categorize a method as pseudo-tested.
For methods with no return value (void-methods), the mutator empties the body of the whole method.
If the resulting mutant survives, the method is categorized as pseudo-tested.
For methods with return values, often multiple mutants have to survive.
For example, for a primitive data type like boolean, two mutants have to survive.
One mutator replaces the method-body with a single return statement that returns \verb|true| and one that only returns \verb|false|.
If both resulting mutants survive, the method is categorized as pseudo-tested.
If only one survives, the method is categorized as \textit{partially-tested}.
Analogously, other mutators that replace the whole method-body with default return values for other data types like integers (e.g. \verb|0| and \verb|1|) or strings (e.g. \verb|""| and \verb|"A"|) are chosen such that, if the mutants survive, the method can be categorized as pseudo-tested.
Complex data types derived from classes, like objects, can be simply set to \verb|null|.
The choice which default values are used depends on the mutation testing tool used \cite{niedermayr2016will,niedermayr_evaluation_2019}.

Compared to traditional mutation testing, extreme mutation testing has the advantage of generating far fewer mutants as the number of methods is significantly lower than the number of instructions that can be mutated.
This reduces both: runtime and time to analyze the mutants.
On the other hand, extreme mutation testing is not as fine-grained as traditional mutation testing to find weaknesses in the test suite.

\section{Method}

\subsection{Research questions}

\subsubsection*{RQ1}
Are the improved execution and analysis times of extreme mutation testing relevant in practice?

\subsubsection*{RQ2}
How do extreme mutation testing results impact the established practice of writing unit tests with code coverage?

\subsubsection*{RQ3}
Which factors prevent extreme mutation testing from industry adoption?

\subsection{Mutation testing}
In the first part of this case study, we have run traditional and extreme mutation testing for a software project that is used in the semiconductor testing industry.
The software is tested by more than 11,000 unit tests that call about 2,000 methods (about 12,500 lines of code).
The software project is written in Java and consists of multiple smaller projects of different age, authors, and coding styles.
For traditional mutation testing, we used {PIT\footnote{http://pitest.org/}}~\cite{coles2016pit} with its default mutation engine Gregor as a mutation testing tool (version 1.4.10).
For extreme mutation testing, we used the same tool but with the Descartes engine\footnote{https://github.com/STAMP-project/pitest-descartes}~\cite{vera2018descartes} (version 1.2.6).
In addition to that, we also measured code coverage with the Java code coverage library JaCoCo\footnote{https://www.jacoco.org/jacoco/} (version 0.8.6) and counted the total lines of code for all methods called by the test suite.

After running mutation tests with different mutators and measuring the code coverage, we additionally labeled the methods by their access modifiers (public, private, etc.) by writing a Python script.
We then manually selected 25 pseudo-tested methods according to the collected data, namely: coverage, number of lines, access modifier, package-, {class-,} and method-name to have a high diversity of methods with different properties.
Afterwards, we killed the extreme mutants that caused the methods to be pseudo-tested, where possible, by enhancing or writing new tests and noted the reason why the method was pseudo-tested.

\subsection{Interview with developers}
The second part of this case study was a semi-structured interview.
We interviewed two developers with more than seven years of development experience who are familiar with the software under test.
None of them used mutation testing beforehand and only one developer heard of it at university.
We structured the interview into four phases.
In the first phase, we asked several questions about the developers' roles in the company and their experiences with unit testing, code coverage, and mutation testing.
Afterwards, we explained traditional and extreme mutation testing and showed the output of the PIT tool.
In the third phase, we opened the software project in an integrated development environment (IDE) and presented pseudo-tested methods to depict the process of how to analyze and strengthen the test suite.
For this session, four classes with eight pseudo-tested methods to discuss were pre-selected.
Five methods had 100\% line coverage and the remaining three had more or equal to 75\% line coverage.
During the enhancement of the test suite, the developers were asked several questions regarding the relevance of the findings and at which stage in development the findings would be most likely fixed.
We also asked whether the developers would have written more or better tests if pseudo-tested lines would be shown as ``uncovered'' in code coverage reports.
The last phase was an open discussion.

\section{Results}

\subsection{Mutation testing results}
\label{quantitative_analysis}
Table~\ref{table:results} shows the results of mutation testing with PIT for 11,424 unit tests that normally finish in 12 seconds.
For traditional mutation testing, we used two sets of mutators that PIT provides.
The ``default'' set consists of seven mutators, whereas the ``all'' set consists of 66 mutators.
All executions were performed on the same machine.

Table~\ref{table:coverage} shows the total number of methods and the number of pseudo-tested methods as well as their proportion.
It also shows the lines that are covered, i.e.,~executed, by the test suite, and the total line counts.
In the analyzed software project, 291 (14\%) of all 2,041 methods are pseudo-tested.
The 291 pseudo-tested methods consist of 1,129 lines of code where 835 lines have coverage.

\begin{table*}[t]
\renewcommand{\arraystretch}{1.3}
\caption{Mutation testing results}
\begin{center}
\begin{tabular}{l r r r r r r}
\hline
\textbf{Type} & \textbf{Score} & \textbf{Killed\,/\,Total} & \textbf{Survived} & \textbf{Mutators} & \textbf{Executed Tests} & \textbf{Time} \\
\hline
Extreme & 85\% & 2,297\,/\,2,706 & 409 & 19 & 597,877 & 13 min \\
Traditional (default) & 73\% & 5,813\,/\,7,989 & 2,176 & 7 & 4,228,169 & 37 min \\
Traditional (all) & 70\% & 55,515\,/\,79,350 & 23,835 & 66 & 34,391,374 & 4 h 1 min \\
\hline
\end{tabular}
\label{table:results}
\end{center}
\end{table*}

We analyzed 25 methods in more depth and found the following three reasons why methods are pseudo-tested:
\begin{itemize}
	\item Weak tests with no assertions (8)
	\item Incomplete tests (3)
	\item Side-effect methods (14)
\end{itemize}
We found eight tests that had no assertions, which caused them to be pseudo-tested.
For example, a class that handled network connections was tested by opening and closing connections, but the connections were never verified to work by, e.g.,~counting the number of open sessions before and after closing.
Instead, these tests were designed in a way that they pass if no failure occurs, which means that no functionality is checked by verifying any output.
Some of these tests could be fixed by simply capturing the output and adding assert statements.
Other tests, like the example that concerns the connections, have required to add new methods to the class under test to get the information like the session count in addition to adding assertions to the test.

\begin{table}[b]
\renewcommand{\arraystretch}{1.3}
\caption{Methods, lines, and coverage}
\begin{center}
\begin{tabular}{l r r r}
\hline
\textbf{Measure} & \textbf{Pseudo-tested} &  \textbf{Total} &  \textbf{Proportion}\\
\hline
Methods & 291 & 2,041 & 14\%\\
Lines (covered) & 835 & 11,189 & 7\%\\
Lines (total) & 1,129 & 12,572 & 9\%\\
\hline
\end{tabular}
\label{table:coverage}
\end{center}
\end{table}

We also found three incomplete tests.
These are tests that are well designed with strong assertions but missed to check certain properties of a class.
For example, a class that creates a table was tested whether it writes the correct data rows, but the table header was not tested, despite being called.
This was easily fixed by adding another assertion for the table header.

The remaining 14 methods are methods that are called during testing but are not subject to be tested and do not have an impact on the test result.
Because of that, we named the last category side-effects.
Typical side-effect methods are, e.g.,~custom methods that are responsible for logging or methods that store meta-data that is never used or verified by the test suite elsewhere.

\subsection{Interview results}

\subsubsection{Main reasons to write unit tests}
\label{reason_unit_tests}
In general, the developers write unit tests to verify that code works as expected, future code changes do not break anything, and to write clean code by using, for example, the test-driven software development approach.
None of the developers write unit tests because of an existing 80\% line-coverage policy in the company.

\subsubsection{Confidence in code coverage}
\label{confidence}
The developers mainly use code coverage to spot regions in the code that have not been tested so far. 
They have some confidence that code coverage is an accurate measure to judge the quality of the test suite when using it for regression testing but are aware of the caveats.
Although one of the main reasons to write unit tests is to verify that the code works as expected, the developers only have very little confidence that code coverage can accurately depict that level of verification.
Code coverage is not seen as an appropriate measure to judge how bug-free the code is.

\subsubsection{Relevance of pseudo-tested methods}
\label{relevance}
The developers see remediating pseudo-tested methods as relevant when they want to improve regression testing and to verify that the code works as expected.
They do not see remediating pseudo-tested methods as relevant to write clean code.

\subsubsection{Pseudo-tested methods in the development process}
\label{development_stage}
When asking the developers at which stage in development the knowledge about pseudo-tested methods would most likely result in improving the test suite, both developers answered at the time when writing unit tests.
The best point in time would be while working with an IDE or when invoking the build on the command line because the developers who write the unit tests have the expert knowledge and could quickly add new tests.
The developers stated that the information would still be useful when a continuous integration pipeline produces a report, but it would strongly depend on the available time and priorities whether they would fix the pseudo-tested method.
They would probably not remediate pseudo-tested methods when spotted during a code review when merging code changes or when receiving the results from a separate quality assurance team because they usually have other priorities at these stages.

\subsubsection{Acting upon pseudo-tested methods}
\label{acting_pseudo_methods}
The developers would have written more or better tests if pseudo-tested methods were highlighted in code coverage reports.
However, the developers stated one exception where they probably would not have enhanced the test suite.
This was the case for a subset of unit tests that tested classes of a particular package but found pseudo-tested methods of classes of a completely different package that was not the target of the unit tests.

\section{Evaluation}

\subsection{Answer to RQ1 -- Relevance of execution and analysis time}
\label{answer_rq1}
Traditional mutation testing with seven default mutators takes almost three times as long (37~min.) as extreme mutation testing (13~min.), despite having only about a third of the number of mutators (see table~\ref{table:results}).
In practice, the execution time requirements strongly depend on in which phase of the development process mutation testing is used.
When analyzing mutation testing results in a separate session, none of the execution times would actually be troublesome because mutation testing would be performed in the background and later analyzed.
However, when trying to perform mutation testing during the development of unit tests, as the interview results suggest (see section~\ref{development_stage}), all of the execution times are too long and thus impractical.
This may be changed when not performing mutation testing for the whole test suite.
For that, extreme mutation testing is more promising due to its lower time requirements.

Table~\ref{table:results} shows that the coarse-grained approach of extreme mutation testing results in having only a fifth of surviving mutants (409) compared to the traditional approach (2,176).
This number gets further lowered to 291 when working on a method-level to check which methods are pseudo-tested (see table~\ref{table:coverage}).
These numbers are more relevant in practice.
If we hypothetically assume that a pseudo-tested method can be remediated in only five minutes, then even remediating 291 pseudo-tested methods would take about 24 hours.
Applying the same calculation to all 2,176 mutants of the traditional approach would result in about seven days of analysis.
Thus, we conclude that the improved execution and analysis times are relevant and noticeable in practice in terms of execution time and workload for the developers.

\subsection{Answer to RQ2 -- Impact on established testing practices}
\label{answer_rq2}
Extreme mutation testing supplements writing unit tests in conjunction with code coverage.
Unit tests are mainly written to verify that the code works as expected and serve as regression tests for future code changes (see sections~\ref{reason_unit_tests}, \ref{confidence}).
Pseudo-tested methods are seen as relevant for improving both (see section~\ref{relevance}).
Line coverage alone does not accurately depict how effective a test suite is.
We found 1,129 lines of code (see table~\ref{table:coverage}) where we could easily introduce new bugs that would remain unnoticed by the given test suite.
In fact, table~\ref{table:coverage} shows that when counting pseudo-tested lines as ``uncovered'', the total line coverage would have to be adjusted from 89\%~to~82\%.

\subsection{Answer to RQ3 -- Industry adoption}
Despite the mentioned advantages in terms of execution and analysis time (see section~\ref{answer_rq1}) and its positive impact on established software testing practices (see section~\ref{answer_rq2}), extreme mutation testing is comparatively young with only a few available implementations~\cite{niedermayr_evaluation_2019} and thus, is not widely known yet.
We noticed that there are two areas to improve to foster industry adoption: tooling and usage.

For tooling, we found that the results of extreme mutation testing would be best represented in code coverage reports by highlighting the pseudo-tested code (see section~\ref{acting_pseudo_methods}).
This is currently not supported by the Descartes engine.
The advantage of that approach is that developers are already familiar with coverage reports, and it is easy to implement.

For usage, further research should investigate how extreme mutation testing has to be improved such that it can be run while writing unit tests.
This is the point in time where issues are most likely fixed (see section~\ref{development_stage}).
Also, we observed that the mutation score is irrelevant when using extreme mutation testing in practice because some methods are not worth fixing (see section~\ref{acting_pseudo_methods}). This was also found in another study~\cite{vera2019comprehensive}.
It would probably be enough to only enhance weak tests with no assertions and incomplete tests (see section~\ref{quantitative_analysis}).
For pseudo-tested side-effect methods, reflecting them in the coverage reports would be sufficient.

\section{Threats to validity}
The presented numbers rely on the correctness of the mutation testing tool PIT.
Some tests of the software project are designed to test parallel execution features.
Other tests rely on static classes and their fields and depend on each other, which influences the test result when executed in parallel or split into independent chunks, which is what PIT does.
Thus, for PIT, we set the number of execution threads to one and tried to remove parallel tests where possible.
Nonetheless, we still found nine methods that belong to a class that included ``parallel'' as part of its name.
However, we ran the extreme and traditional mutation tests with default mutators both three times and always got the same results.

Further, we only interviewed two experienced developers who mostly agreed in their opinions.
This does not necessarily mean that the results are well generalizable and that other developers would rate the relevance of pseudo-tested methods or how to use extreme mutation testing the same way they did.

\section{Conclusions}
We have shown that extreme mutation testing is supplemental to current software testing practices.
It can help to write better unit tests and addresses some shortcomings that code coverage has.
Compared to the traditional approach, the faster execution time is well noticeable in practice and provides a good starting point for further research.
For result presentation, we consider it to be best integrated into existing coverage reports because it is easy to implement and it is a format that is already known by the developers.
Future research should investigate how to apply extreme mutation testing while writing unit tests such that the results are available fast enough but still meaningful.
Lastly, future research should also analyze the severity and number of issues that remain undiscovered by the extreme approach, which trades accuracy for speed gains.

\bibliographystyle{IEEEtran}
\bibliography{IEEEabrv,references.bib}

\begin{thebibliography}{1}
\providecommand{\url}[1]{#1}
\csname url@samestyle\endcsname
\providecommand{\newblock}{\relax}
\providecommand{\bibinfo}[2]{#2}
\providecommand{\BIBentrySTDinterwordspacing}{\spaceskip=0pt\relax}
\providecommand{\BIBentryALTinterwordstretchfactor}{4}
\providecommand{\BIBentryALTinterwordspacing}{\spaceskip=\fontdimen2\font plus
\BIBentryALTinterwordstretchfactor\fontdimen3\font minus
  \fontdimen4\font\relax}
\providecommand{\BIBforeignlanguage}[2]{{%
\expandafter\ifx\csname l@#1\endcsname\relax
\typeout{** WARNING: IEEEtran.bst: No hyphenation pattern has been}%
\typeout{** loaded for the language `#1'. Using the pattern for}%
\typeout{** the default language instead.}%
\else
\language=\csname l@#1\endcsname
\fi
#2}}
\providecommand{\BIBdecl}{\relax}
\BIBdecl

\bibitem{jia2010analysis}
Y.~Jia and M.~Harman, ``An analysis and survey of the development of mutation
  testing,'' \emph{IEEE transactions on software engineering}, vol.~37, no.~5,
  pp. 649--678, 2010.

\bibitem{papadakis2019mutation}
M.~Papadakis, M.~Kintis, J.~Zhang, Y.~Jia, Y.~Le~Traon, and M.~Harman,
  ``Mutation testing advances: an analysis and survey,'' in \emph{Advances in
  Computers}.\hskip 1em plus 0.5em minus 0.4em\relax Elsevier, 2019, vol. 112,
  pp. 275--378.

\bibitem{niedermayr2016will}
R.~Niedermayr, E.~Juergens, and S.~Wagner, ``Will my tests tell me if i break
  this code?'' in \emph{2016 IEEE/ACM International Workshop on Continuous
  Software Evolution and Delivery (CSED)}.\hskip 1em plus 0.5em minus
  0.4em\relax IEEE, 2016, pp. 23--29.

\bibitem{niedermayr_evaluation_2019}
\BIBentryALTinterwordspacing
R.~Niedermayr, ``Evaluation and improvement of automated software test
  suites,'' Ph.D. dissertation, University of Stuttgart, 2019. [Online].
  Available: \url{http://dx.doi.org/10.18419/opus-10640}
\BIBentrySTDinterwordspacing

\bibitem{coles2016pit}
H.~Coles, T.~Laurent, C.~Henard, M.~Papadakis, and A.~Ventresque, ``Pit: a
  practical mutation testing tool for java,'' in \emph{Proceedings of the 25th
  International Symposium on Software Testing and Analysis}, 2016, pp.
  449--452.

\bibitem{vera2018descartes}
O.~L. Vera-P{\'e}rez, M.~Monperrus, and B.~Baudry, ``Descartes: a pitest engine
  to detect pseudo-tested methods: tool demonstration,'' in \emph{2018 33rd
  IEEE/ACM International Conference on Automated Software Engineering
  (ASE)}.\hskip 1em plus 0.5em minus 0.4em\relax IEEE, 2018, pp. 908--911.

\bibitem{vera2019comprehensive}
O.~L. Vera-P{\'e}rez, B.~Danglot, M.~Monperrus, and B.~Baudry, ``A
  comprehensive study of pseudo-tested methods,'' \emph{Empirical Software
  Engineering}, vol.~24, no.~3, pp. 1195--1225, 2019.

\end{thebibliography}
\end{document}